\documentclass[a4paper,fleqn,useAMS,usenatbib]{mnras}

\pdfoutput=1

\usepackage{mathptmx}

\usepackage[T1]{fontenc}
\usepackage{ae,aecompl}

\usepackage[dvips]{graphicx}
\usepackage{amsmath}
\usepackage{amsfonts}
\usepackage{amssymb}
\usepackage{comment}
\usepackage[dvipsnames]{xcolor}
\usepackage[%
        ]{hyperref}
        
\hypersetup{
	colorlinks=true,	
	urlcolor=MidnightBlue,		
	pdfpagelabels=true,
	hypertexnames=true,
	plainpages=false,
	naturalnames=true,
	pdftitle={A machine learns to predict the stability of circumbinary planets},    
	pdfauthor={Lam \& Kipping},     
	linkcolor=WildStrawberry,          
	citecolor=ForestGreen,        
}

\newcommand{\python}{{\tt Python}}
\newcommand{\keras}{{\tt keras}}
\newcommand{\rebound}{{\tt REBOUND}}
\newcommand{\batman}{{\tt batman}}
\newcommand{\reboundlink}{\href{http://github.com/hannorein/rebound}{this URL}}

\title[A machine learns to predict the stability of circumbinary planets]{
A machine learns to predict the stability of circumbinary planets
}
\author[Lam \& Kipping]{Christopher Lam$^{1}$\thanks{E-mail:
\href{mailto:cl3425@columbia.edu}{cl3425@columbia.edu}} and David Kipping$^{1}$\\
$^{1}$Dept. of Astronomy, Columbia University, 550 W 120th Street, New York NY 10027}

\date{Accepted . Received ; in original form }

\pubyear{2017}

\begin{document}
\label{firstpage}
\pagerange{\pageref{firstpage}--\pageref{lastpage}}
\maketitle

\begin{abstract}
Long-period circumbinary planets appear to be as common as those
orbiting single stars and have been found to frequently have orbital radii
just beyond the critical distance for dynamical stability. Assessing the
stability is typically done either through N-body simulations or using the stability criterion first considered by Dvorak and later developed by Holman and Wiegert: a second-order polynomial
calibrated to broadly match numerical simulations. However, the polynomial is
unable to capture islands of instability introduced by mean motion resonances,
causing the accuracy of the criterion to approach that of a random coin-toss
when close to the boundary. We show how a deep neural network (DNN) trained on
N-body simulations generated with \rebound\ is able to significantly improve
stability predictions for circumbinary planets on initially coplanar, circular orbits.
Specifically, we find that the accuracy of our DNN never drops below 86\%, even
when tightly surrounding the boundary of instability, and is fast enough to be practical
for on-the-fly calls during likelihood evaluations typical of modern Bayesian
inference. Our binary classifier DNN is made publicly available at
https://github.com/CoolWorlds/orbital-stability.
\end{abstract}

\begin{keywords}
circumbinary planets --- machine learning --- deep neural networks
\end{keywords}

\section{Introduction}
\label{sec:intro}

The majority of exoplanets discovered to date reside in systems which comprise
of three or more components (see
\href{https://exoplanetarchive.ipac.caltech.edu}{NEA}; \citealt{akeson:2013}).
In such a regime, the analytic solution of the two body problem is not directly
applicable and, in general, the orbits must be computed numerically
\citep{murray:2010}. Further, whilst the two body problem guarantees
dynamically stable orbits, three-or-more body systems can become unstable over
the span of many orbits, leading to ejections or collisions of one or more
members \citep{murray:1999}.

When studying a specific three-or-more body exoplanet system, a common task is
to regress a model describing the observations in question that, in general,
must compute the orbital motion of the system's components. In many cases, the
temporal baseline under analysis is much shorter than the expected timescale
for significant departures away from purely Keplerian-like motion, which
greatly expedites the analysis (e.g. see \citealt{luna:2011} in the case of
exomoons). However, although the observed baseline is short, the system is
presumably old and thus must have survived a much greater number of orbits
without becoming dynamically unstable. This latter point means that,
in principle, regressions of three-or-more body systems should check the
resulting dynamical stability of each trial set of parameters. Due to the
high computational cost of numerical stability simulations, this is rarely
plausible given modern computational facilities, particularly when performing
Bayesian regressions which typically explore millions of possible states.

Conventionally, the approach to this problem is to simply ignore it,
test the stability of a small subset of fair samples \citep{lissauer:2011}, or
use an approximate analytic metric to test stability, such as the Hill
criterion \citep{deck:2013}. Recently, there has been interest in adapting
the latter approach from using approximate analytic formulae to test stability
at each trial, to calling a pre-trained machine learning algorithm capable
of rapidly assessing stability. Specifically, \citet{tamayo:2016} demonstrated
the plausibility of this approach in application to compact multi-planet
systems. In this work, we consider another important case study to be that of
circumbinary planets.

Long-period ($\lesssim$300\,d), giant ($R_P>6$\,$R_{\oplus}$) circumbinary
planets appear to be as common as those orbiting single stars
\citep{armstrong:2014}. No short period circumbinary planets are presently
known\footnote{Kepler-47b is the current shortest period circumbinary
planet with $P\simeq50$\,days \citep{orosz:2012}.}, which is expected given the
unstable nature of such orbits due to perturbations from the binary itself
\citep{holman:1999}. Since the highly successful transit method is heavily
biased against long-period planets \citep{sandford:2016,gongjie:2016}, this
means that only a handful of circumbinary planets are presently known. However,
the continual accumulation of photometric data sets on overlapping fields, such
as TESS \citep{ricker:2015}, NGTS \citep{wheatley:2013} and PLATO
\citep{rauer:2014}, means that more discoveries can be expected of these
remarkable exoplanet systems.

The most common approach to evaluating the stability of circumbinary planets is the analytic formula developed from \citet{dvorak} by \citet{holman:1999}, who regressed a second-order polynomial to a suite of dynamical simulations. However, as noted in that work, islands of instability existed which were not captured by this simple formalism. Since many circumbinary planets appear to reside
close to the \citet{holman:1999} critical orbital radius \citep{welsh:2014}, there is a pointed need to test the long-term stability of trials explored during a Bayesian regression, to enable tighter, more physically-sound parameter inference. Following the approach of \citet{tamayo:2016}, we turn to machine learning to improve upon the \citet{holman:1999} stability criterion with the goal of capturing the islands of instability noted by the authors.

In Section~\ref{sec:training}, we first describe how we generated a large suite
of training data using numerical simulations. In Section~\ref{sec:learning},
we describe our learning algorithm using a deep neural network. In
Section~\ref{sec:performance}, we explore the performance of our approach,
particularly in comparison to the existing method of \citet{holman:1999}.
Finally, we discuss possible applications of our method and future areas for
development in Section~\ref{sec:discussion}. 

\section{Training Data}
\label{sec:training}

\subsection{Numerical simulations with \rebound}
\label{sub:rebound}

In order to obtain the training data for our deep network, we decided to use
the \rebound\ N-body integrator first developed by \citet{rebound}. Since first
publication, \rebound\ has evolved to include IAS15, a non-symplectic
integrator with adaptive time-stepping \citep{ias15}. We use IAS15 as our integrator of choice. Within the \rebound\ environment, we set up a circumbinary planet system with
two stars orbiting each other with a semi-major axis of $a_{\mathrm{bin}}$ and
orbital eccentricity $e_{\mathrm{bin}}$. The planet is chosen to orbit the
binary on an initially circular, coplanar orbit with semi-major axis $a_P$.
We ran $10^6$ \rebound\ experiments each comprising of ten simulations, with
each simulation lasting ten thousand binary periods. Experiments were aborted
if the planet met our ejection criterion described later in
Section~\ref{sub:stabcheck}. 

For each simulation, we uniformly sampled the mass ratio, $\mu$, from
values between 0 and 0.5, where

\begin{align}
\mu \equiv \frac{m_a}{m_a+m_b},
\end{align}

where $m_a \leq m_b$,
and the binary eccentricity, $e_{\mathrm{bin}}$, from values between 0 and
0.99. We then sampled the initial semi-major axis of the planet uniformly from
an envelope of $\pm33\%$ surrounding the empirically derived \citet{holman:1999}
stability criterion for a circumbinary planet\footnote{We initially experimented
with 10\% to 50\% envelopes, but found the 33\% envelope was sufficient to
encompass the islands of instability, yet compact enough to focus on the areas of
interest.}, which is given by

\begin{align} 
a_{\mathrm{HW99}} &= (1.60\pm0.04) + (5.10\pm0.05) e \nonumber\\
 & \qquad + (-2.22\pm0.11) e^2 + (4.12\pm0.09) \mu \nonumber\\
 & \qquad + (-4.27\pm0.17) e \mu + (-5.09\pm0.11) \mu^2 \nonumber\\
 & \qquad + (4.61\pm0.36) e^2 \mu^2 .
\label{eq1} 
\end{align}

Here, $a_{\mathrm{HW99}}$ is the critical semi-major axis given in units of binary separation, since gravity's lack of scale allows us to use dimensionless units. Orbital elements are Jacobi elements, so positions, velocities, and other orbital elements are centered about the binary center of mass.

To take phase into account, for every sample, ten initial binary phases were
drawn uniformly from [0,2$\pi$), while keeping the planet phase at zero. If any one of these ten simulations led to an
unstable orbit, the grid point was labelled as ``unstable''; otherwise, it
was ``stable''. In total, this meant that ten million unique \rebound\
simulations were executed. The stability label of each of the $10^6$ grid
points was saved and used for training our machine learning algorithm later.

\subsection{Stability criterion}
\label{sub:stabcheck}

Systems can become unstable by the planet being either ejected from the system,
or colliding with one of the stars. The former instance can be tested by
evaluating whether the radial component of the planet's velocity, $v_r$,
exceeds the escape velocity of the binary pair, $v_{\mathrm{esc}}$, indicating 
that the planet will escape to infinity. Rather than formally test for collisions,
which requires setting physical radii, we label cases where the planet's orbit
passes interior to that of the binary orbit as being ``unstable''.

\begin{figure}
\begin{center}
\includegraphics[width=8.4cm]{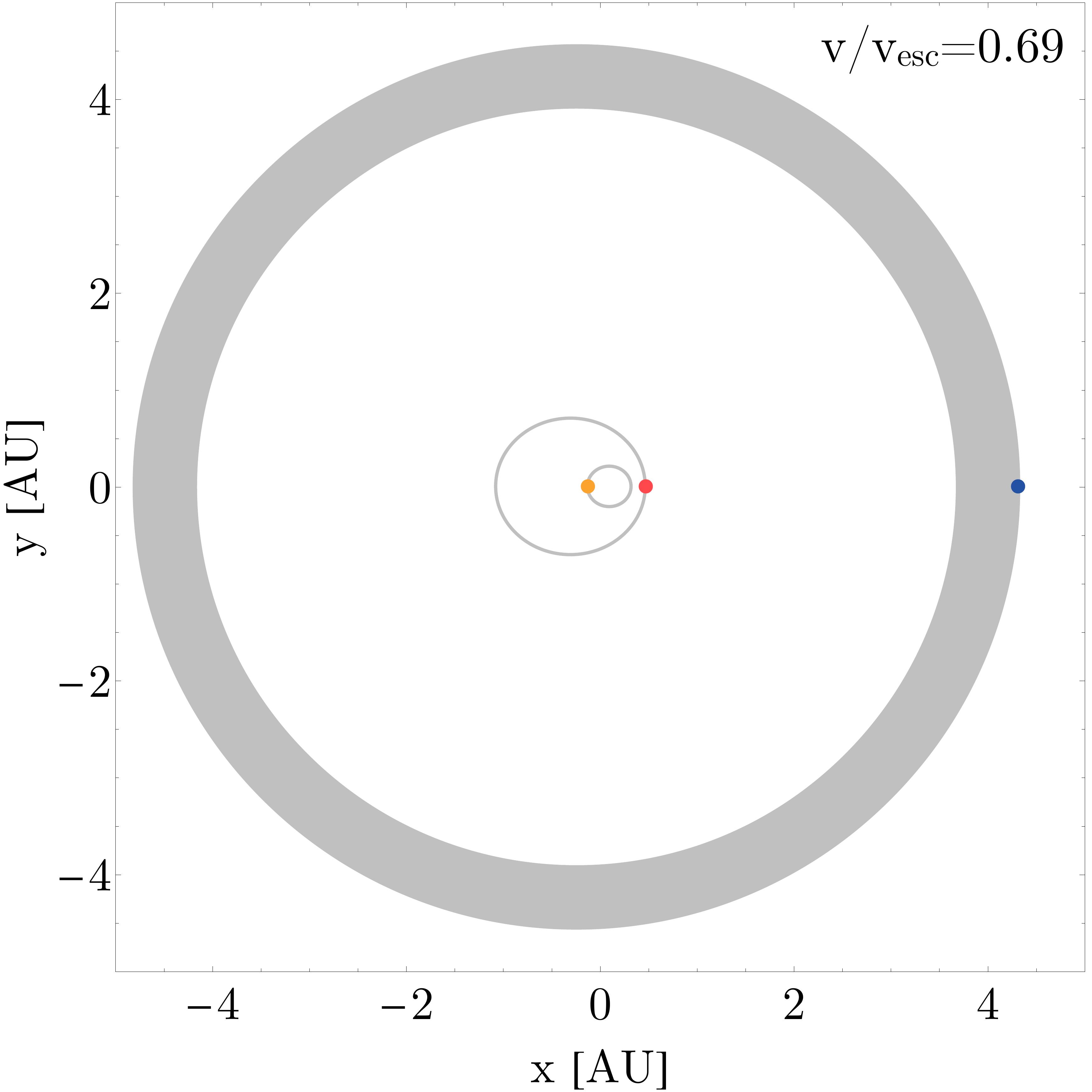}
\caption{
Example of a stable system after ten thousand binary periods.
Initial parameters were $e_{\mathrm{bin}}$ of 0.4, $\mu$ of 0.5, and $a_P$ of
20\% greater than the \citet{holman:1999} criterion for the given
\{$e_{\mathrm{bin}}$, $\mu$\}. 
}
\label{fig:stable}
\end{center}
\end{figure}

\begin{figure}
\begin{center}
\includegraphics[width=8.4cm]{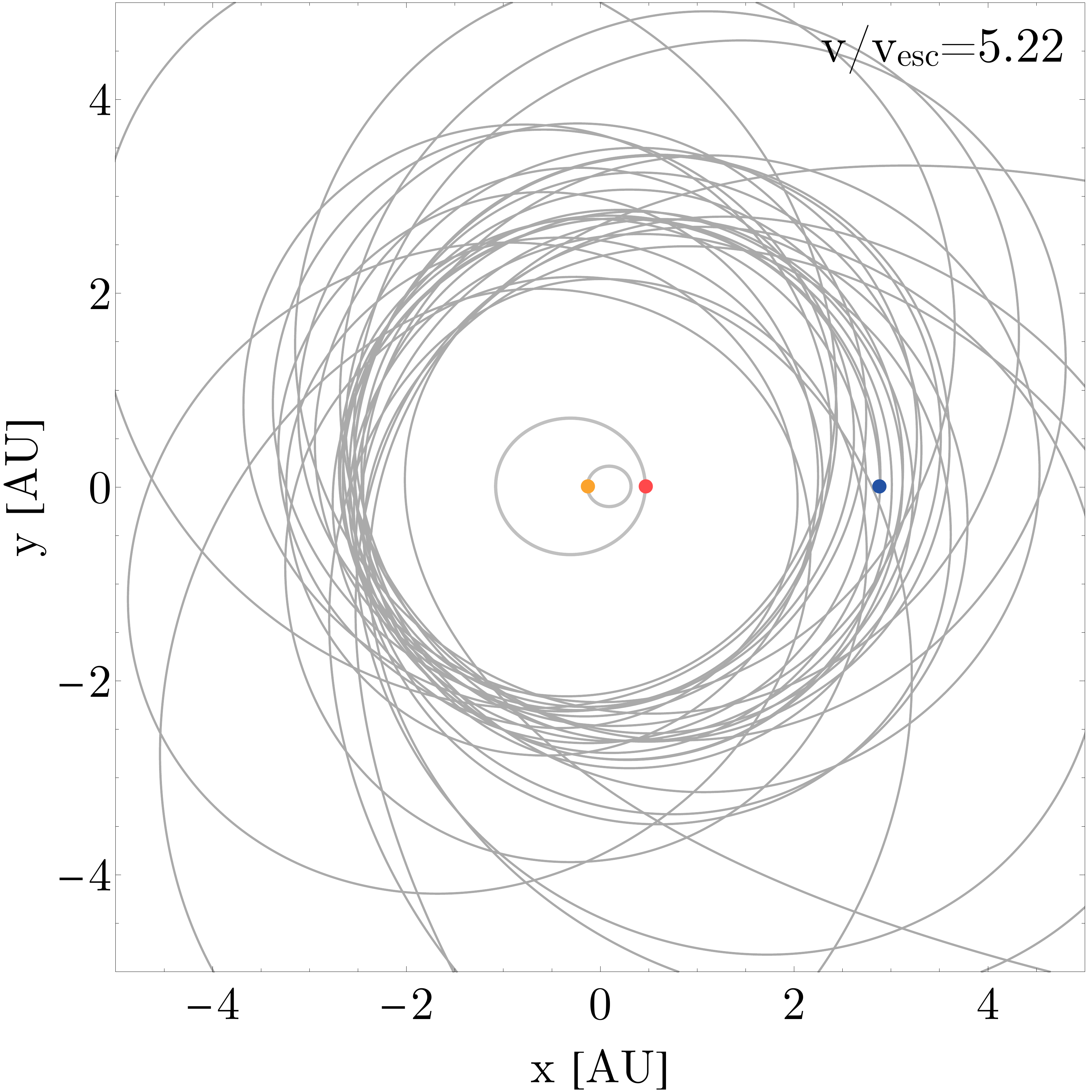}
\caption{
Example of an unstable system after a few dozen binary periods.
Initial parameters are $e_{\mathrm{bin}}$ of 0.4, $\mu$ of 0.5, and $a_P$ of
20\% smaller than the \citet{holman:1999} criterion for the given
\{$e_{\mathrm{bin}}$, $\mu$\}.
}
\label{fig:unstable}
\end{center}
\end{figure}

In Figures~\ref{fig:stable} \& \ref{fig:unstable}, we show two typical examples of
a random system which was labelled as ``stable'' and ``unstable''.

\subsection{Islands of instability}
\label{sub:islands}

As can be seen later in Figure~\ref{fig:performance}, the loci of stable
points in $\{e_{\mathrm{bin}},a_P/a_{\mathrm{bin}}\}$ parameter space
is described not just by a broad curve, which represents the
\citet{holman:1999} critical radius, but also numerous ``islands of
instability'', to quote the original authors. A machine learning algorithm
capable of capturing these islands would lead to a significant improvement
in the accuracy of predicting long-term stability versus the simple
polynomial form of \citet{holman:1999}. These islands are driven by mean
motion resonances, and occur at regular intervals in period ratios going as
2:1, 3:1, 4:1, etc.

Due to these resonances, naively using the \citet{holman:1999} stability
criterion sometimes misclassifies stability up to a certain distance from
$a_{HW99}$. We conducted a quick check for the minimum-sized envelope
containing all such misclassifications, finding that it was sufficient to
automatically classify areas further than $120\%$ of the $a_{HW99}$ as stable
and those interior to $80\%$ of $a_{HW99}$ as unstable. The network was
therefore trained on the 600826 samples that fell within the $20\%$ envelope.

To accomodate calls to our algorithm which exceed the 20\% region, 
we simply use an if-statement to assign ``stable'' if above and ``unstable''
if below.

\section{Supervised Learning}
\label{sec:learning}

\subsection{Deep Neural Networks (DNNs)}
\label{sub:DNN}

Whilst \citet{tamayo:2016} predicted the orbital stability of planets in
compact multi-planet systems using the {\tt XGBoost v0.6} algorithm
\citep{xgboost:2016}, we elected instead to employ deep neural networks
(DNNs), which are comprised of layers through which data is propagated in order
to make classifications and predictions. DNNs have been demonstrated to be
powerful tools in predictive applications, including already within the field
of exoplanetary science (for example see \citealt{transit:2017}, which also
includes a brief pedagogical explanation of neural networks).

In this work, we used the \keras\ neural network \python\ library rather than construct our own implementation, largely because the DNN required was deeper and more
intricate than that used in our earlier work of \citet{transit:2017}.

\subsection{Network architecture}
\label{sub:architecture}

DNNs make predictions by taking a set of $N$-dimensional inputs and passing the
data through ``hidden'' layers of neurons, in which nonlinear activation
functions transform the data. Hidden layers are linked by synapse-like weights
that transform the data linearly. The output of the network - here, a binary
classification of stable or unstable - is compared with the true results, and
the error calculated by some loss function. Our learning algorithm
employs back-propagation, which is the chain rule applied to the loss
function with respect to the parameter weights. In this manner, the learning
algorithm optimizes the loss function by gradient descent.  This DNN
implementation is typically referred to as a multi-layer perceptron (MLP).

Several key decisions factored into the neural net's architecture. These
include the number of hidden layers, the number of neurons in these hidden
layers, the activation functions, the loss function, and dropout.   

We used a fairly standard design, comprised of six layers of 48 hidden
neurons each, using relu (rectifier linear unit) and sigmoid activation
functions, \keras's {\tt binary\_crossentropy} loss function, and the {\tt
rmsprop} optimizer. The relu activation function, which lives in the hidden layers, is defined as 

\begin{align}
f(x) \equiv \mathrm{max}(0,x) .
\end{align}

The sigmoid activation function, which lives in the single-neuron output
layer, outputs a probability from which classes can be produced and is defined
as

\begin{align}
f(x) = \frac{1}{1+e^{-x}}.
\end{align}

The {\tt rmsprop} optimizer normalizes a moving average of the root mean
squared gradients (described in detail within the package documentation
at \href{http://climin.readthedocs.io/en/latest/rmsprop.html}{this URL}).

Dropout is a technique to avoid overfitting, an issue endemic to models as complex as DNNs. By setting the dropout rate of a hidden layer, one can adjust the probability at which any hidden neuron and its associated weights drop out of the training set for any particular pass through the neural net. Over many passes, different thinned versions of the DNN are trained on the data, resulting in neurons that depend less on each other for learning, which is also known as a decrease in ``co-adapting''. Compared to the classic L2 and L1 regularization methods, dropout has been shown to guard against overfitting better for neural networks  (\citealt{dropout:2014}).

\subsection{Feature selection}
\label{sub:features}

Since we sought to compare our model against the previous literature, namely
the \citet{holman:1999} criterion, we initially only trained our DNN on
the same three variables of that work: $e_{\mathrm{bin}}$, $\mu$ and
$(a_P/a_{\mathrm{bin}})$. However, we found that even after experimenting with
the network architecture, training sets and learning algorithms, the DNN was
unable to reliably recover the islands of instability.

The inability of our initial DNNs to capture the islands motivated us to
add an additional fourth training feature. Specifically, since the islands
are located at mean motion resonances, we decided to add a new feature
describing how far away from one of these resonances a trial semi-major axis
is. We refer to this term as the ``resonance proxy'', $\epsilon$, in what
follows and define it mathematically as

\newcommand\floor[1]{\lfloor#1\rfloor}
\begin{align}
\epsilon \equiv \frac{\zeta - \floor{\zeta}}{2},
\end{align}

where $\floor{\zeta}$ is the floor of the ratio of semi-major axes converted to orbital period
space via Newton's version of Kepler's Third Law, such that

\begin{align}
\zeta(a_p,a_{\mathrm{bin}}) = (a_p/a_{\mathrm{bin}})^{3/2}
\end{align}

After introducing this fourth feature, we found dramatically improved
performance of the DNN in terms of capturing the islands of instability,
and thus adopted the feature in what follows.

\subsection{Iterative learning}

When training a DNN, one generally aims to pass the training data through
the network iteratively until the weights of the neural net are satisfactorily
tuned. To achieve this, we increased the number of training rounds, measured in
``epochs'', until the loss on the validation set stopped significantly
decreasing. An ``epoch'' is simply a forward pass through the network and the
subsequent back-propagation for all training examples. The number of epochs,
then, is the amount of training done before validating the model. 

We trained the DNN over several different choices for the maximum number of
epochs, ranging from 10 to 250. In general, we found that even after 10 rounds
the weights were well-tuned, which we assessed by comparing the
cross-validation precision-recall curves resulting from each and described
later in Section~\ref{sub:PreRec}. The learning was relatively fast, taking under 12 seconds per epoch on a MacBook Air. In our final implementation, we elected to use the 100 epoch trained network, whose performance is described quantitatively in Section~\ref{sec:performance}.

\section{Network Performance}
\label{sec:performance}

\subsection{Capturing the islands of instability}

\begin{figure*}
\begin{center}
\includegraphics[width=17.0cm]{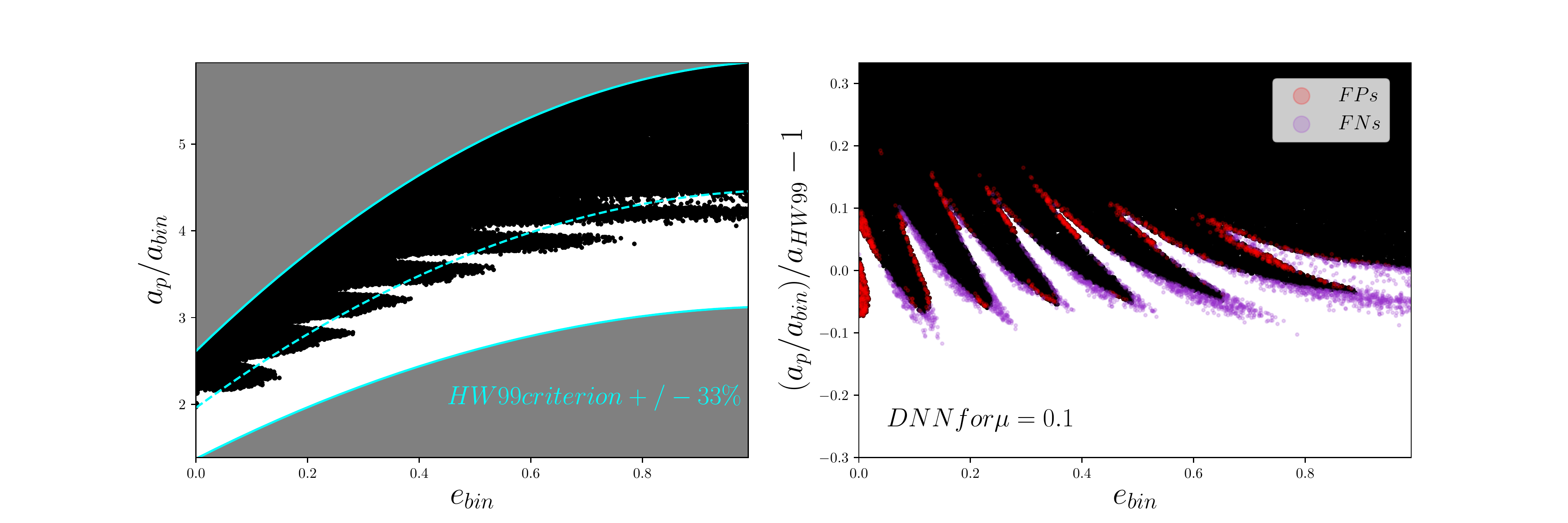}
\caption{
\textit{Left panel:} $10^5$ randomly generated, initially circular, coplanar
circumbinary planets, where the binary's eccentricity is varied
($x$-axis) and the planet's relative semi-major axis is varied
($y$-axis). Islands of dynamical instability, as first noted by
\citet{holman:1999} [HW99], appear at mean motion resonances yet
are not captured by the HW99 polynomial (blue dashed).
\textit{Right panel:} Stability predictions from our neural net
for the same $10^5$ points versus \rebound\ simulations. The net
was trained on different data, yet is able to capture the
islands of instability with a low rate of false-positives (FPs)
and false-negatives (FNs). To better visualize the islands of instability, we re-parameterize, treating the Holman-Wiegert criterion as a baseline by dividing the semi-major axis ratio by the criterion and subtracting by one.
}
\label{fig:performance}
\end{center}
\end{figure*}

We find that our DNN, as desired, is able to accurately re-produce the
islands of instability first noticed by \citet{holman:1999}, but not
captured by their second-order polynomial function. This can be
seen in Figure~\ref{fig:performance} where we show a slice through
parameter space for a fixed $\mu$ set to 0.1. Here, we generate $10^5$
random new samples in $\{e_{\mathrm{bin}},(a_P/a_{\mathrm{bin}})\}$
parameter space within the $\pm33$\% envelope used earlier but hold
$\mu$ constant at $0.1$ each time, thereby generating a new
validation set across a 2D slice. This presentation enables us to
easily visualize the islands of instability (left panel) yet also see
that in our re-parameterized domain, we are able to accurately
predict these islands using our DNN (right panel).

\subsection{Execution time}

With the learning finished and weights pre-set in the network, any DNN is 
generally expected to be very fast in making forward predictions. Indeed, on a MacBook Air, the call time for making
predictions on $10^5$ validation examples is approximately 29 seconds, or
about 3 ten-thousandths of a second per validation example. Such a computation time is
on the same order of magnitude as typical forward models for
describing exoplanet transit observations - for example, the popular transit
package \batman\ requires around 0.4\,milli-seconds per call with non-linear
limb darkening\footnote{Although \batman\ does not natively handle circumbinary
planets, which would likely increase this computation time.}
\citep{kreidberg:2015}.

Accordingly, our DNN is fast enough to be practical for on-the-fly Bayesian
samplers that potentially need to call the routine many millions of times.
Such an approach allows for the computing of a posterior probability that accounts
for dynamical stability, where the simplest implementation truncates
unstable samples to a zero density.

\subsection{Accuracy, Precision, and Recall}
\label{sub:PreRec}
\begin{figure*}
\begin{center}
\includegraphics[width=17.0cm]{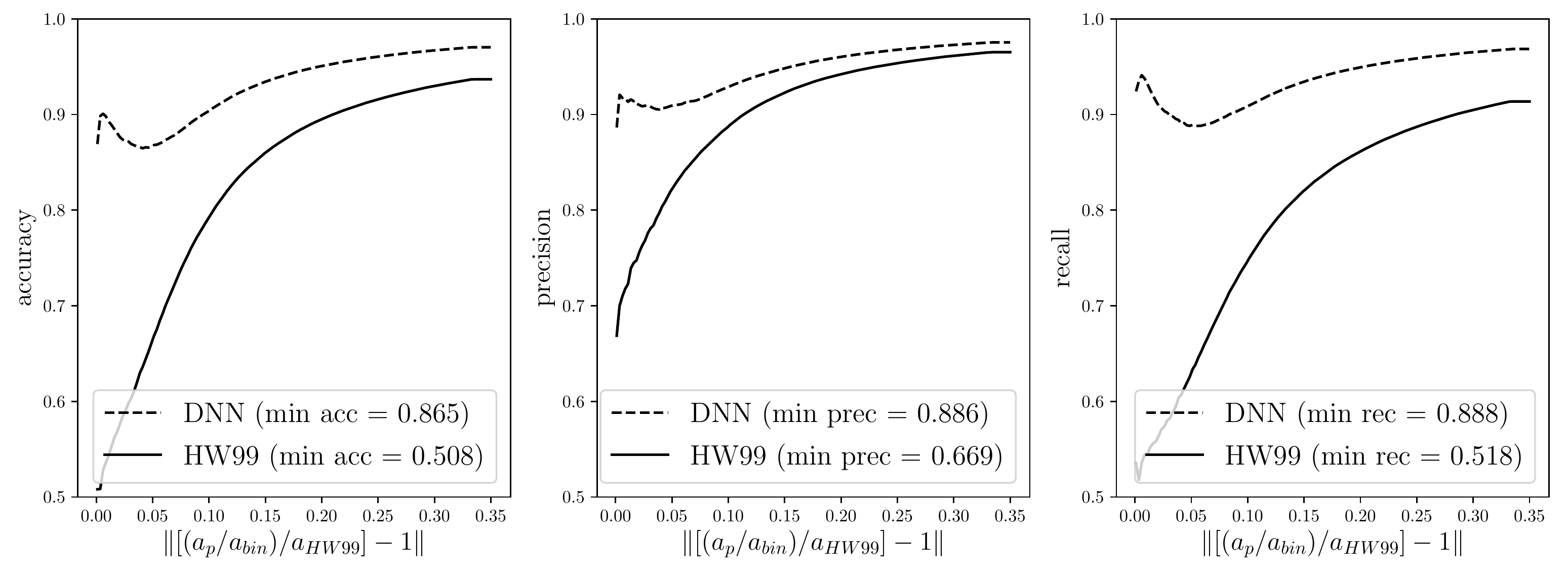}
\caption{
\textit{Left panel:} accuracy
\textit{Middle panel:} precision
\textit{Right panel:} recall
\hspace{\textwidth}
Defining the full envelope as the set of all samples where the absolute value of the planet's semi-major axis is less than 33\%, we observe that our DNN only marginally outperforms Holman-Wiegert on accuracy, precision, and recall. We express the envelopes by re-parameterizing the semi-major axis ratio of planet to binary, dividing it by the polynomial from \citet{holman:1999} to normalize, and subtracting by one to center the islands of instability at zero. Moving left along the x-axis, we shrink the envelope, increasing the dominance of the islands - which are the most difficult to capture - in the validation set. Overall, while Holman-Wiegert converges towards a random Bernoulli process, our DNN persists near 90\% for accuracy, precision, and recall.
}
\label{fig:accprerec}
\end{center}
\end{figure*}

When computing statistics by which to evaluate our model, it was important that we not simply
apply our DNN directly on the trained data but rather apply it to a previously
unseen set, known as a validation set. Accordingly, we generated a new
random batch of $10^5$ \rebound\ simulations within the same bounding region
as that originally trained on. We then applied our DNN and tallied
the true positive (TP), false positive (FP), true negative (TN), and false negative (FN) rates to generate the accuracy, precision, and recall plots shown in
Figure~\ref{fig:accprerec}. As evident from the figure, our DNN indeed
out-performs the \citet{holman:1999} criterion, which supports our earlier comparison in Figure~\ref{fig:performance}. This is
to be expected given that our DNN has a far larger number of weights than the
coefficients describing the \citet{holman:1999} polynomial, allowing it to better express phenomena as complex as the islands of instability.

We compared our DNN with the \citet{holman:1999} criterion using three statistics based on TP, FP, TN, and FN. One basic and useful way to quantify the performance of a binary classifier is accuracy, which is the number of correct classifications as a fraction of total cases. It is defined as

\begin{align}
\mathrm{accuracy} \equiv \frac{ \mathrm{TP} + \mathrm{TN} }{ \mathrm{TP} + \mathrm{TN} + \mathrm{FP} + \mathrm{FN} }.
\end{align}

Taking the entire suite of validation samples, we find that the accuracy of
the \citet{holman:1999} criterion is 93.7\%, whereas the DNN achieves 97.1\%
accuracy over the same region. However, the outer edges of this region are
deep into stable and unstable territories, where both approaches have a relatively
easy job of picking out the correct answers. To more rigorously evaluate the models, we filter our validation set down to a sub-sample closely surrounding
the \citet{holman:1999} boundary. We vary the envelope sequentially in small
steps, starting from a very tight envelope and gradually expanding out to the
far regions, and in each case compute the accuracy of the two approaches.

The results are shown in the left panel of Figure~\ref{fig:accprerec}, where one can see that the DNN
out-performs the \citet{holman:1999} criterion across the entire range of
parameter space considered. As expected, one can see the two functions converge
as the envelope grows towards infinity, where the border-line cases become
an increasingly negligible fraction of the overall sample. We find that the
\citet{holman:1999} criterion approaches a pure random Bernoulli process
for very tight envelopes, which is generally to be expected. In contrast,
the DNN performs well even at the boundary and reaches its lowest accuracy
of 86.5\% close to the boundary. These results allow us to predict that our DNN
will be able to deliver $\geq86.5$\% accuracy for predicting the stability of
circumbinary planets on initially circular, coplanar orbits.

Two other statistics that are common in classifier evaluation are precision, which is defined as the fraction of positive
predictions that are actually true, and recall, which is the fraction of true
examples that were predicted to be positive. Mathematically, they can be
defined as follows:

\begin{align}
\mathrm{precision} \equiv \frac{\mathrm{TP}}{\mathrm{TP} + \mathrm{FP}}
\end{align}

\begin{align}
\mathrm{recall} \equiv \frac{\mathrm{TP}}{\mathrm{TP} + \mathrm{FN}}
\end{align}

\citet{tamayo:2016} evaluated the performance of their network using the classic precision-recall framework, which is typically well-suited for comparing classifiers whose output probabilities can be thresholded. Whereas the thresholds of the Lissauer-family models they evaluated against are varying Hill-sphere separations, the \citet{holman:1999} criterion only predicts stability based on a single threshold: whether or not a planet's semi-major axis is above or below the curve of the polynomial. Instead of plotting precision against recall, we consequently plotted each against incrementally smaller envelope sizes, as we did with accuracy. While our DNN marginally outperforms the \citet{holman:1999} criterion for the full envelope, with 97.5\% versus 96.5\% precision and 96.8\% versus 91.4\% recall, it demonstrates a slower drop in precision and recall as the envelope shrinks. In the smaller-envelope regions of the islands of instability, the performance difference widens to greater than 20\% for precision and 35\% for recall. 

\section{Discussion}
\label{sec:discussion}

In this work, we have demonstrated that a deep neural network (DNN) is able to predict the dynamical stability of circumbinary planets on initially coplanar,
circular orbits with greater accuracy, precision, and recall than the previous
state-of-the-art: a polynomial function presented in \citet{holman:1999}. Our
paper builds upon the recent work of \citet{tamayo:2016}, who showed that
machine learning techniques are well-suited for predicting the stability of
multi-planet systems (indeed that was our original applied system too when
starting this work). It also provides another powerful demonstration that
machine learning is well-suited for binary classification problems in
exoplanetary science, such as our previous work using DNNs to predict whether
a known planetary system has an additional outer transiting
planet \citep{transit:2017}.

By measuring the accuracy, precision, and recall of our DNN predictions on validation sets, we find that our DNN consistently
out-performs the \citet{holman:1999} stability criterion, especially in its ability to capture the islands of instability caused by mean motion resonances described in the original paper of \citet{holman:1999}. Over the entire region
simulated, which spans $\pm33$\% of the \citet{holman:1999} criterion, we find that our DNN modestly outperforms the criterion in accuracy, precision, and recall. However, as we decrease the envelope size from $\pm33$\% towards 0\% so that the islands of instability dominate the validation set, both classifiers face a tougher challenge, with the
\citet{holman:1999} criterion approaching a random Bernouilli process and
our DNN hovering slightly below 90\% accuracy and precision, and slightly above 90\% recall. Although our simulations are averaged
over random initial phases, it may be that improving metrics such as accuracy beyond this
point will be fundamentally limited by the chaotic nature of quasi-stable
orbits on the boundary. Despite experimentation with our network configuration,
we were unable to make any improvements to the performance beyond the version
described in this work.

We would like to highlight that our work by no means devalues the
polynomial-based approach of \citet{holman:1999}. Indeed, our training was
greatly aided by this previous seminal work, which allowed us to immediately
focus in on the region of interest. Further, such analytic formulae provide
opportunities for re-parameterization (although we did not pursue this
directly on our trained data set), which could also benefit machine learning
algorithms.

Our work has assumed a circular, coplanar initial orbit for the planet, as well
as test mass particles. For any plausible parameters, the stability boundary extends at least a few tens of stellar radii, so the effects of tides and GR can be ignored, since they are negligible compared to the libration timescale inside the mean motion resonances. Exploring eccentric, initially non-coplanar orbits would seem to be of particular
importance for future work, since several known circumbinary planets indeed have
small eccentricities. Another possible relevant addition to the feature set is the initial binary phase, which we had accounted for by sampling uniformly as described in Section~\ref{sub:rebound}. We hope that our work, particularly the realization of
using resonance-proxy as an extra feature, may aid and guide those attempting to
build more sophisticated classifiers in the future. 

The main motivation and obvious application for this work is to apply the
classifier during the act of Bayesian regression of circumbinary planet
systems. Often these algorithms attempt millions or even billions of trial
solutions, which may appear reasonable solutions in terms
of likelihood yet dive into regions of parameter space where stability
is questionable. In such cases, our classifier could be practically used
on-the-fly to evaluate the stability of each trial, and thus assign a
stability probability which could be multiplied by the likelihood. The
simplest way to do this would be to use the binary classification output
directly (i.e. the probability is either zero or unity).

We finish by highlighting that machine learning techniques are well-suited
for problems such as dynamics, where we have in hand the ability to
generate essentially arbitrarily large training sets. Deep networks
in particular require very large training sets but are able to map
non-linear, pathological functions which occur in dynamical problems.
	
\section*{Acknowledgments}

This work has been supported by the Columbia University Center for Career Education's
Work Exemption Program and made use of Columbia's Habanero shared research
computing facility.
Simulations in this paper made use of the \rebound\ code which can be
downloaded freely at \reboundlink.
This research has made use of the NASA Exoplanet Archive, which is operated by
the California Institute of Technology, under contract with the National
Aeronautics and Space Administration under the Exoplanet Exploration Program.
We thank Daniel Tamayo, Hanno Rein, Gongjie Li, and the Cool Worlds Lab team for helpful comments and
conversations in preparing this manuscript.


%

\bsp
\label{lastpage}
\end{document}